\documentclass[12pt]{article}

\usepackage[margin=1in]{geometry}
\usepackage{setspace}
\usepackage{graphicx}
\usepackage{amsmath}
\usepackage{amssymb}
\usepackage[super]{cite}

\usepackage{orcidlink}

\usepackage{authblk}

\begin{document}
	
	\title{Light Propagation through Space-Time Non-Markovian Random Media}
	
	\author[1,2]{Chaoran Wang~\orcidlink{0009-0009-6585-7199}\thanks{chaoranwang@siom.ac.cn}}
	\author[1,2]{Jinquan Qi~\orcidlink{0009-0004-7183-4210}}
	\author[1]{Shuang Liu~\orcidlink{0000-0003-2504-2349}}
	\author[1]{Chenjin Deng~\orcidlink{0000-0002-1403-0177}}
	\author[1,2,3]{Shensheng Han~\orcidlink{0000-0003-1689-5876}\thanks{sshan@mail.shcnc.ac.cn}}
	
	\affil[1]{Aerospace Laser Technology and System Department, Shanghai Institute of Optics and Fine Mechanics, Chinese Academy of Sciences, Shanghai 201800, China}
	\affil[2]{Center of Materials Science and Optoelectronics Engineering, University of Chinese Academy of Sciences, Beijing 100049, China}
	\affil[3]{Hangzhou Institute for Advanced Study, University of Chinese Academy of Sciences, Hangzhou 310024, Zhejiang, China}
	
	\date{} 
	\maketitle
	
	\begin{abstract}
		Here, we introduce a stochastic partial differential equation (SPDE) formulation driven by temporally correlated noise to describe light propagation beyond the standard Markov approximation. By representing the squared refractive index fluctuations as a random field with explicit long-range temporal correlations, we demonstrate that the propagation dynamics map exactly onto the hyperbolic Anderson model. This rigorous mapping enables the derivation of new quantitative scaling relations that connect the environment's non-Markovian memory effects to the statistical properties of the emergent light field. We experimentally validate these analytical predictions in an outdoor atmospheric environment, confirming the memory-dependent statistical signatures of the propagated light. Our results establish a precise physical foundation for understanding memory-driven wave phenomena, providing crucial insights for free-space optical communication, remote sensing, and coherent imaging.
	\end{abstract}
	
	\section{Introduction}
	The problem of light propagation through space-time random media (e.g., the atmosphere and ocean \cite{he2017SpaceTimeCorrelationsDynamic,rotter2017LightFieldsComplex,weiss1961WavePropagationWave,vynck2023LightCorrelatedDisordered}) has attracted significant interest in diverse fields, including optical communication and imaging \cite{gozzard2022UltrastableFreeSpaceLaser,zhang2010CorrelatedImagingAtmospheric,parry1979DistributionsAtmosphericPropagation,popoff2010MeasuringTransmissionMatrixa}, quantum information and sensing \cite{capraro2012ImpactTurbulenceLong,guo2017EntanglementdistillationAttackContinuousvariable}, adaptive optics \cite{davies2012AdaptiveOpticsAstronomy,horst2023TbitLinerateSatellite}, photon fluid \cite{situ2020DynamicsBerezinskiiKosterlitz}, among others.
	
	The traditional theoretical frameworks for investigating this problem generally rely on the Markov approximation, which assumes that refractive index fluctuations are delta-correlated along the direction of propagation \cite{tatarskii1980IIIStrongFluctuations,belenkii1980CoherenceFieldLaser}. While methods such as the Rytov approximation \cite{tataraskii1969LightPropagationMedium,andrews2005LaserBeamPropagation} and the higher-order moment equations \cite{brown1971SecondMomentWave,jakeman1978SignificanceDistributionsScattering,garnier2016FourthMomentAnalysisWave} have proven effective under specific conditions, their inherent memoryless assumption neglects the cumulative effects of long-range correlations along the propagation path. Consequently, fully resolving propagation dynamics becomes challenging for these methods in realistic media where the scale of inhomogeneities is comparable to the path length or where temporal memory effects play a critical role.
	
	To address the complexity of realistic media, numerical simulation methodologies have been widely adopted, most notably the multiple phase screen model \cite{booker1985ComparisonExtendedmediumPhasescreen,andrews1997PropagationGaussianbeamWave} and Monte Carlo methods \cite{ishimaru1989DiffusionLightTurbid,prahl1989MonteCarloModel}. Although these approaches effectively model light intensity distributions \cite{zhang2010CorrelatedImagingAtmospheric,song2023PathSamplingIntegration} and geometric beam deformations \cite{paterson2005AtmosphericTurbulenceOrbital,huang2014EvolutionBehaviorGaussian}, they typically treat propagation as a sequence of independent scattering events. This discrete layering enforces a quasi-Markovian structure, thereby discarding continuous non-Markovian characteristics, such as the temporal memory effects and spatial long-range dependence prevalent in turbulent environments \cite{caruso2014QuantumChannelsMemory,bulgac2024QuantumTurbulenceSuperfluidity}. Consequently, a comprehensive model capable of intrinsically characterizing light transmission within non-Markovian space-time random media remains imperative.
	
	We transcend the limitations of memoryless approximations by deriving a stochastic partial differential equation (SPDE) through an operator separation technique. Unlike the standard parabolic equation governed by the Markov approximation, our derived equation corresponds formally to the hyperbolic Anderson model in SPDE theory. This framework naturally accommodates colored noise and temporal memory, extending the conventional parabolic approximation by explicitly accounting for the finite correlation time of the media. Based on this derivation and recent theoretical advances \cite{nualart2006MalliavinCalculusRelated,dalang1999ExtendingMartingaleMeasure,dalang2009StochasticWaveEquation,balan2012StochasticWaveEquation,balan2022ExactAsymptoticsStochastica,balan2025GaussianFluctuationsWave}, we analytically map the media's non-Markovian stochastic descriptors to the statistical properties of the light field. Furthermore, outdoor experiments were conducted to verify the validity of these analytical results in a real-world environment exhibiting strong memory effects.
	
	\section{Theory}The electromagnetic field $u(t,\boldsymbol{x})$ propagating in a space-time uniform medium satisfies the partial differential equation (PDE) \cite{evans2010PartialDifferentialEquations}:
	$	\mathcal{L}_{c_{a}} u(t,\boldsymbol{x}) = 0,$
	with $\mathcal{L}_{c_{a}}:=\frac{n^{2}_{a}}{c^{2}}\frac{\partial^{2}}{\partial t^{2}}-\Delta_{\boldsymbol{x}}$, where
	$\Delta_{\boldsymbol{x}}=\frac{\partial^{2}}{\partial x^{2}}+\frac{\partial^{2}}{\partial y^{2}}+\frac{\partial^{2}}{\partial z^{2}} $ is the Laplacian operator in three dimensions, $n_{a}$ is the refractive index of the uniform medium, and $c$ is the wave speed of light in a vacuum. Similar to the potential function applied in time-dependent scattering theory \cite{vynck2023LightCorrelatedDisordered}, we decompose the square of the refractive index, thereby splitting the operator into two parts:
	\begin{align}
		\frac{n^{2}(t,\boldsymbol{x})}{c^{2}}\frac{\partial^{2}}{\partial t^{2}}-\Delta_{\boldsymbol{x}}&=\frac{\mathbb{E}[n^{2}(t,\boldsymbol{x})]+\mu(t,\boldsymbol{x})}{c^{2}}\frac{\partial^{2}}{\partial t^{2}}-\Delta_{\boldsymbol{x}}\notag\\
		&=\mathcal{L}_{c_{n}}+\frac{\mu(t,\boldsymbol{x})}{c^{2}}\frac{\partial^{2}}{\partial t^{2}}, \label{w1}
	\end{align}
	here $\mu(t,\boldsymbol{x})$ is a space-time random field representing the fluctuation of the square of the media's refractive index, and $\mathcal{L}_{c_{n}}=\frac{\mathbb{E}[n^{2}(t,\boldsymbol{x})]}{c^{2}}\frac{\partial^{2}}{\partial t^{2}}-\Delta_{\boldsymbol{x}}$ is a deterministic second-order partial differential operator. Notably, $\mathbb{E}[\cdot]$ denotes ensemble averaging and $\left\langle \cdot \right\rangle $ denotes averaging over time. With the decomposition in Eq.\eqref{w1}, one can rewrite the wave equation as:
	\begin{align}
		&\mathcal{L}_{c_{n}}u(t,\boldsymbol{x})+\frac{\mu(t,\boldsymbol{x})}{c^{2}}\frac{\partial^{2}u(t,\boldsymbol{x})}{\partial t^{2}} = 0.
		\label{w2}
	\end{align}
	Because $\mu(t,\boldsymbol{x})$ is a random field, it is challenging to obtain a solution for Eq.\eqref{w2} directly using Sturm-Liouville theory \cite{evans2010PartialDifferentialEquations}.
	
	Within the specific context considered here, there are no strong nonlinear effects during the long-distance propagation process, and the time derivative of the phase fluctuations of light fields induced by random media is significantly lower than the optical angular frequency. Thus, Eq.\eqref{w2} transforms into the following form:
	\begin{align}
		\mathcal{L}_{c_{n}}u(t,\boldsymbol{x}) &= {W}(t,\boldsymbol{x})u(t,\boldsymbol{x}),
		\label{w3}
	\end{align}
	where $
	{W}(t,\boldsymbol{x}):=\frac{\mu(t,\boldsymbol{x})\omega^{2}}{c^{2}}$, and $\omega $ is the angular frequency of the light field.
	
	The SPDE structure defined in Eq.\eqref{w3} is recognized in the mathematical literature as the hyperbolic Anderson model \cite{dalang2009IntermittencyPropertiesHyperbolic}. Central to solving this model is the accurate characterization of the random field $W(t, \boldsymbol{x})$, which must account for the complex space-time correlations of the media. It is well-established that random media—particularly fully developed turbulence—exhibit multifractal scaling properties within the spatial domain \cite{leonardis2013IdentificationIntermittentMultifractal,xu2006MultifractalDimensionLagrangian}. Specifically, the autocorrelation function of the scalar field in such media follows a power-law spectrum characterized by $\vert\boldsymbol{r}\vert^{-\gamma}$ \cite{poujade2006RayleighTaylorTurbulenceNothing,ristorcelli2004RayleighTaylorTurbulence}. Here $\vert \boldsymbol{r}\vert$ represents the Euclidean distance between two points, while $\gamma$ denotes the spatial power-law indicator, often referred to as the Extended Self-Similarity (ESS) parameter. The quantitative characterization of distinct turbulent regimes is thus contingent upon the specific values assigned to $\gamma$ \cite{stribling1995OpticalPropagationNonKolmogorov,ristorcelli2004RayleighTaylorTurbulence}. 
	
	Crucially, from the perspective of time dynamics, realistic random media are predominantly non-Markovian in most practical scenarios \cite{balkovsky2001IntermittentDistributionInertial}. Within this framework, the conventional Markovian assumption, which presupposes memoryless fluctuations, is relegated to a simplified special case \cite{perez2004FractionalBrownianMotion,biagini2008StochasticCalculusFractional}. By contrast, our approach recognizes that realistic random media are predominantly non-Markovian, characterized by intrinsic temporal correlations that the standard Markovian limit fails to capture. To rigorously incorporate these intrinsic non-Markovian temporal correlations, we employ fractional Brownian motion (fBm), $B^{H}_{t}$, to characterize the long-range dependence of the media in the time domain. By synthesizing these temporal characteristics with the spatial multifractal scaling, we model the total covariance function of the space-time random media as:
	\begin{align}
		\mathbb{E}[{W}(t,\boldsymbol{x}),{W}(s,\boldsymbol{y})]=C_{H}(t^{2H}+s^{2H}-\vert t-s\vert^{2H})\vert\boldsymbol{x}-\boldsymbol{y}\vert^{-\alpha},
		\label{w4}
	\end{align}
	where $s,t,\alpha \in \mathbb{R}_{+}$, $\boldsymbol{x}, \boldsymbol{y}\in \mathbb{R}^{3}$, and $\vert \boldsymbol{x}-\boldsymbol{y} \vert \geq l_{0}$. Here, $H \in (0,1)$ represents the Hurst index, a fundamental parameter that characterizes the non-Markovian long-range dependence of the stochastic process:
	\begin{itemize}
		\item $H > 1/2$: persistence, characterized by a positive autocorrelation.
		\item $H < 1/2$: anti-persistence, characterized by a negative autocorrelation.
		\item $H = 1/2$: the memoryless Markovian limit, corresponding to standard Brownian motion \cite{biagini2008StochasticCalculusFractional}.
	\end{itemize}
	The term $\vert\boldsymbol{x}-\boldsymbol{y}\vert^{-\alpha}$ corresponds to the Riesz kernel function utilized in the hyperbolic Anderson model. While the parameter $\alpha$ can be determined through experimental measurement, the restriction $\vert \boldsymbol{x}-\boldsymbol{y} \vert \geq l_{0}$ is introduced to prevent divergence at the origin. Mathematically, however, it is sufficient for the covariance function to satisfy the Dalang condition to ensure the existence of a valid solution \cite{dalang1999ExtendingMartingaleMeasure,dalang2009StochasticWaveEquation}.
	
	When the covariance function conforms to Eq.\eqref{w4}, there exists a wild solution $u_{w}(t,\boldsymbol{x})$ to Eq.\eqref{w2}:
	\begin{align}
		&u_{w}(t,\boldsymbol{x})= u_{n}(t,\boldsymbol{x})+ \int_{0}^{t}\int_{\mathbb{R}^{3}}G(t-t_{1},\boldsymbol{x}-\boldsymbol{x}_{1})W(dt_{1},d\boldsymbol{x}_{1})\notag\\
		&+\int_{0}^{t}\int_{\mathbb{R}^{3}}G(t-t_{1},\boldsymbol{x}-\boldsymbol{x}_{1})
		\int_{0}^{t_{1}}\int_{\mathbb{R}^{3}}G(t_{1}-t_{2},\boldsymbol{x}_{1}-\boldsymbol{x}_{2}) W(dt_{1},d\boldsymbol{x}_{1}) W(dt_{2},d\boldsymbol{x}_{2})+\cdots. \label{w5}
	\end{align}
	Here, $t>t_{1}>t_{2}>\cdots>0$, $G(\cdot,\cdot)$ is the Green's function of $\mathcal{L}_{c_{n}}u(t,\boldsymbol{x})=0 $, and $u_{n}(t,\boldsymbol{x})$ is the deterministic solution of $\mathcal{L}_{c_{n}}u(t,\boldsymbol{x})=0$ with the same boundary condition and initial value condition \cite{evans2010PartialDifferentialEquations,balan2012StochasticWaveEquation}. Thus, $u_{w}(t,\boldsymbol{x})$ constitutes a random field in the distributional sense \cite{m.balan2017HyperbolicAndersonModel}. To further analyze the properties of this solution, we apply the Malliavin calculus method, rewriting Eq.\eqref{w5} as \cite{nualart2006MalliavinCalculusRelated,nourdin2012NormalApproximationsMalliavin}:
	\begin{align}
		u_{w}(t,\boldsymbol{x}) =  \sum^{\infty}_{n= 0} J_{n}(f_{n}(\cdot,t,\boldsymbol{x})),
	\end{align}
	where $J_{n}$ denotes the $n$-th Wiener-Chaos expansion \cite{balan2012StochasticWaveEquation,nourdin2012NormalApproximationsMalliavin}, and $f_{n}$ is the symmetrized distribution in the Schwartz Space $\mathcal{S}^{\prime}(\mathbb{R}^{nd})$:
	\begin{align}
		f_{n}(\cdot ,t,\boldsymbol{x})&=\frac{1}{n!}\sum_{\pi \in S_{n}}G(t-t_{\pi(n)},\boldsymbol{x}-\boldsymbol{x}_{\pi(n)}) G(t_{\pi(n)}-t_{\pi(n-1)},\boldsymbol{x}_{\pi(n)}-\boldsymbol{x}_{\pi(n-1)})\cdots\notag\\
		& G(t_{\pi(2)}-t_{\pi(1)},\boldsymbol{x}_{\pi(2)}-\boldsymbol{x}_{\pi(1)})1_{0<t_{\pi(1)}<\cdots<t_{\pi(n)}<t},
	\end{align}
	where $\pi$ runs over all permutations of $1,2,\cdots,n$, and here $\cdot$ denotes the missing $n$ variables $(t_{\pi(1)},\boldsymbol{x}_{\pi(1)}),\cdots,(t_{\pi(n)},\boldsymbol{x}_{\pi(n)})$. More precisely, the Wiener-Chaos expansions of $W(t,\boldsymbol{x})$ collectively generate the direct sum space $\oplus_{n=0}^{\infty}\mathcal{H}_{n}$, which is isomorphic to a Hilbert space equipped with the $L^{2}$ norm \cite{balan2012StochasticWaveEquation,nourdin2012NormalApproximationsMalliavin}, and $J_{n}$ is a linear operator from $\mathcal{H}^{\oplus n}$ onto $\mathcal{H}_{n}$.
	
	For the variance of the intensity $I(t,\boldsymbol{x}) = \vert u_{w}(t,\boldsymbol{x})\vert ^{2}$ of the transmitted light field, an explicit analytical expansion can be derived as:
	\begin{align}
		\mathrm{Var}\left[I(t,\boldsymbol{x})\right] = \sum_{j=1}^{\infty}\frac{(-1)^{j+1}}{j!}\mathbb{E}\left[\left(D^{j}I(t,\boldsymbol{x})\right)^{2}\right], \label{w9}
	\end{align} 
	where $D$ is the Malliavin differential operator associated with $W(t,\boldsymbol{x})$, and the superscript $j$ indicates that the Malliavin differential operator acts $j$ times. 
	
	We define the strength of the intensity fluctuation as the scintillation index $ \sigma ^{2}$ \cite{berman2009InfluencePhasediffuserDynamics,parry1979DistributionsAtmosphericPropagation}:
	\begin{align}
		\sigma^{2} := \left\langle\Delta I^{2}\right\rangle/\left\langle I\right\rangle^{2}=(\langle I^{2}\rangle/\langle I\rangle^{2})-1.
	\end{align}

	By applying Fubini's theorem and the dominated convergence theorem \cite{klebaner1998IntroductionStochasticCalculus,biagini2008StochasticCalculusFractional}, it is straightforward to deduce from Eq.\eqref{w9} that the scintillation index after propagating a distance $L$ is strictly bounded above by a finite constant:
	\begin{align}
		\sigma^{2}\big\vert^{(\infty)}_{L}&\simeq \sum_{j=1 }^{\infty}\frac{(-1)^{j+1}}{j!}\int_{S(l_{0})}^{S(L)}j\vert \boldsymbol{r} \vert ^{-\alpha+2j-2} d\boldsymbol{r} \notag\\
		&=\int_{S(l_{0})}^{S(L)}\sum_{j=1 }^{\infty}\frac{(-1)^{j-1}}{(j-1)!}\vert \boldsymbol{r}\vert^{2(j-1)}\vert \boldsymbol{r} \vert ^{-\alpha} d\boldsymbol{r} \notag\\
		&=\int_{S(l_{0})}^{S(L)}e^{-\vert \boldsymbol{r}\vert^{2}}\vert \boldsymbol{r} \vert ^{-\alpha} d\boldsymbol{r}\notag\\
		&\rightarrow \frac{\pi^{3/2}}{2\Gamma(3/2)} \cdot \Gamma \left( \frac{3-\alpha}{2} \right)<\infty,\;\mathrm{as}\; l_{0}\rightarrow 0 ,L\rightarrow \infty. \label{w11}
	\end{align}
	The upper bound indicates that the model proposed here is able to explain the scintillation saturation phenomenon observed in experiments \cite{berman2009InfluencePhasediffuserDynamics}.

	When we take the first $N+1$ $(2\leq N<\infty)$ items in the sequence, the infinite series in formula Eq.\eqref{w9} is truncated. By further setting $H = 1/2$, the discrete version can be formally obtained by invoking the Lagrange integral mean-value theorem (for the explicit discretization procedure, see Supplementary Information) \cite{SeeSupplementalMaterialb}:
	\begin{align}
		\sigma^{2}\big\vert^{(N)}_{L}
		&\simeq  \sum_{k=1}^{N}\int_{S(l_{k-1})}^{S(l_{k})} \sum_{j=1}^{k}\frac{(-1)^{j-1}}{(j-1)!}\vert \boldsymbol{r}\vert^{2(j-1)} \vert \boldsymbol{r} \vert ^{-\alpha} d\boldsymbol{r}\notag\\
		&\simeq \sum_{k=1}^{N} V_{k} \sum_{j=1}^{k}\frac{(-1)^{j-1}}{(j-1)!}\vert \boldsymbol{r}_{k}\vert^{2(j-1)} \vert \boldsymbol{r}_{k} \vert ^{-\alpha}. \label{w20}
	\end{align}
	
	Here, $V_{k}:=\int_{S(l_{k-1})}^{S(l_{k})}\mathbf{1}d\boldsymbol{r}$ denotes the volume measure of the integration region over $S(l_{k-1})$ and $S(l_{k})$, and $\{ l_{0},l_{1},\cdots,l_{N-1},L,\vert l_{k-1}<\vert \boldsymbol{r}_{k}\vert<l_{k}\} $ is a sequence with elements spaced at the same interval of $(L-l_{0})/N-1$. Furthermore, by regarding each item of Eq.\eqref{w20} as describing the effect of the corresponding phase screen on the scintillation, and taking the limit $N\rightarrow\infty$, we can use the squeeze theorem to explain why the multiple phase screen method can approximately account for the saturation of scintillation when $N\gg 1$ \cite{zhang2010CorrelatedImagingAtmospheric,booker1985ComparisonExtendedmediumPhasescreen}.
	
	Knowing all the higher-order moments of the random field $ u_{w}(t,\boldsymbol{x}) $, we can obtain a complete description of its overall probability distribution through its generating function. 
	
	Utilizing the above relationship for lower-order moments versus higher-order moments, for all integers $p,q\geq 2$, at the spatial coordinate $\boldsymbol{x}_{0}$, the following exact relationships hold \cite{balan2022ExactAsymptoticsStochastica}:
	\begin{align}
		\lim_{t\rightarrow\infty}\log \mathbb{E}u_{w}^{p}(t,\boldsymbol{x}) &=  p\left(\frac{p-1}{2}\right)^{\frac{1}{2-\alpha}}C_{1}(\alpha,H)t^{\frac{4H-\alpha}{2-\alpha}},\label{w12}\\
		\dfrac{\langle I^{q}\rangle}{\langle I \rangle^{q}} \bigg\vert_{\boldsymbol{x}_{0}}
		&=\left(\dfrac{\langle I^{2}\rangle}{\langle I \rangle^{2}}\bigg\vert_{\boldsymbol{x}_{0}}\right)^{\frac{q(2q-1)^{\frac{1}{2-\alpha}}-q}{2\cdot3^{\frac{1}{2-\alpha}}-2}}, \label{w13}
	\end{align}
	where $C_{1}(\alpha,H)$ is a deterministic constant dictated by $H$ and $\alpha$, whose exact analytical form is provided in the Supplementary Information \cite{SeeSupplementalMaterialb}. Formula Eq.\eqref{w12} describes the asymptotic behavior of the solution when the temporal variable $t$ in the hyperbolic Anderson model tends to infinity. As the spatial power-law index $\alpha$ appears explicitly in expression Eq.\eqref{w13}, the space-time characteristics of the media are reflected in the distribution of the propagated light field.
	
	Subsequently, we analyze the temporal properties of the solution. At a given spatial point, by the isometry property, we have \cite{balan2010StochasticWaveEquationa}:
	\begin{align}
		\mathbb{E}u_{w}(t,\boldsymbol{x})u_{w}(s,\boldsymbol{x}) &= C_{H}\int_{\mathbb{R}^{3}}\int_{0}^{t}\int_{0}^{s}dvdw\vert v-w\vert^{2H-2}\mathcal{F}G(v-t,\boldsymbol{\xi})\overline{\mathcal{F}G(w-s,\boldsymbol{\xi})}\Lambda(d\boldsymbol{\xi}), \label{whurt}
	\end{align}
	where $\Lambda(\cdot)$ denotes the associated Lebesgue measure of $\vert \boldsymbol{x}\vert ^{-\alpha}$. It is observed that when the Hurst index exceeds 1/2, the covariance function obeys the following identity \cite{balan2010StochasticWaveEquationa}:
	\begin{align}
		&t^{2H}+s^{2H}-\vert t-s\vert^{2H}\equiv2H(2H-1)\int_{0}^{t}\int_{0}^{s}dvdw\vert v-w\vert^{2H-2}. \label{iwx}
	\end{align}
	
	Eq.\eqref{whurt} and Eq.\eqref{iwx} demonstrate that when the Hurst exponent satisfies $H > 1/2$, the random field $W(t, \boldsymbol{x})$ exhibits long-range positive temporal dependence. Under this condition, the light field $u_w(t, \boldsymbol{x})$ at any fixed spatial location $\boldsymbol{x}_0$ likewise displays persistent temporal correlation. Conversely, for $H < 1/2$, a negatively correlated random field induces analogous anti-persistent behavior in the resulting light field \cite{poujade2006RayleighTaylorTurbulenceNothing,biagini2008StochasticCalculusFractional}. This dual dependence is also corroborated by Eq.\eqref{w12} and Eq.\eqref{w13}, wherein the identity $(4H - \alpha)/(2 - \alpha) \equiv 1$ holds exactly at $H = 1/2$, marking the transition between correlation regimes.
	
	In practical optical systems, the measured field always represents an averaging of the detected field over the detector area, which can be mathematically described as $F_{R}(t)=\int_{B(R^{2},z_{0})}u_{w}(x,y,z_{0},t)dxdy$, where $B(R^{2},z_{0})$ denotes a disk of radius $R$ at propagation distance $z_{0}$, and $R$ is controlled by the entrance pupil of the optical system.

	By rigorously applying the functional central limit theorem \cite{balan2025GaussianFluctuationsWave}, we establish the following upper bound for the convergence:
	\begin{align}
		d_{\mathrm{sKL}}\left(Z_{R}(t),\mathcal{N}_{t} (0,1)\right) \leq C_{2} R^{-\frac{\alpha}{2}}, \label{w15}
	\end{align}
	here $d_{\mathrm{sKL}}$ denotes the symmetrized Kullback-Leibler (sKL) divergence between two probability density functions (see Supplementary Information for its precise definition in this context) \cite{SeeSupplementalMaterialb}, $Z_R(t) := \{F_R(t) - \langle F_R(t) \rangle\} / \sqrt{\mathrm{Var}(F_R(t))}$, $\mathcal{N}_t(0, 1)$ represents a continuous-time sampling from the standard complex Gaussian distribution with zero mean and unit amplitude variance, and $C_2$ is a constant. Eq.\eqref{w15} implies that even when $H \neq 1/2$, in the limit of a large spatial integration region, the normalized process $Z_R(t)$ converges in distribution to the standard complex Gaussian distribution, with a Hurst index of $1/2$. Moreover, the convergence rate is at least polynomial with exponent $-\alpha/2$.
	
	Notably, when $H \neq 1/2$, the light field after propagation exhibits non-Markovian fluctuations under small-aperture detection. Such colored noise is known to be more detrimental to optical communication systems than white noise. The upper bound established in Eq.\eqref{w15} provides a theoretical foundation for the empirical observation that aperture averaging or spatial diversity reception can effectively mitigate degradation induced by random media in long-distance optical propagation \cite{andrews2005LaserBeamPropagation,khalighi2009FadingReductionAperture}.
	\begin{figure}
		\centering
		\includegraphics[height=7cm]{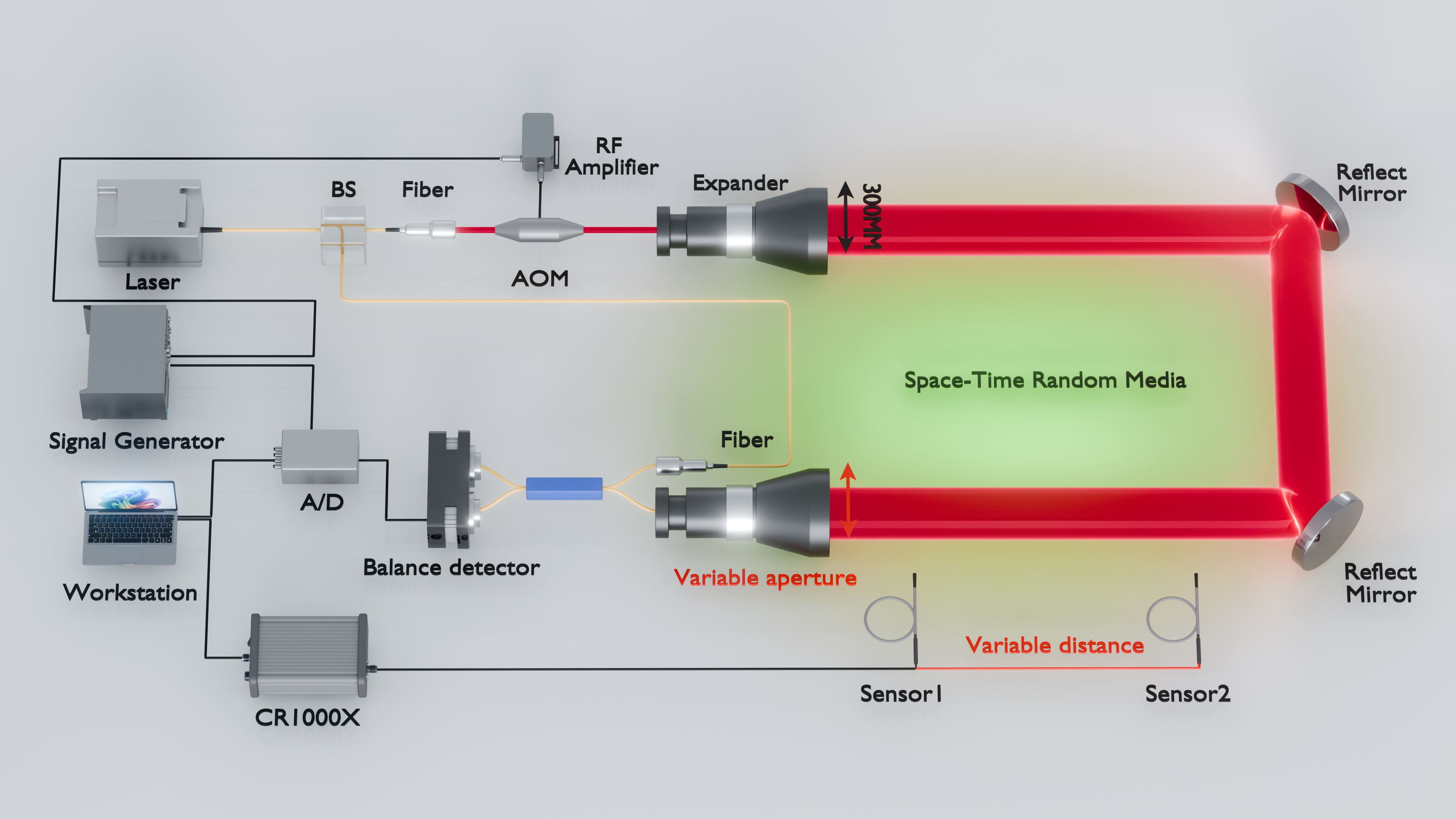}
		\caption{The schematic diagram depicts the two-point temperature difference measurement device and the large aperture coherent detection system used in this experiment.}
		\label{F1}
	\end{figure}
	
\section{Experiment and Results}

To rigorously validate the proposed theoretical framework, we designed and implemented a comprehensive outdoor experimental platform. This platform integrates two synchronized modalities: a highly sensitive micrometeorological array to probe the local refractive index fluctuations $\mu(t, \boldsymbol{x})$ across distinct spatial points, and a large-aperture heterodyne coherent detection system to capture the precise amplitude and phase dynamics of a laser field propagating through the realistic, non-Markovian atmospheric channel.

The physical parameters of the optical system were strictly tailored to isolate the environment's memory effects. Specifically, a highly stabilized continuous-wave laser featuring an ultra-narrow intrinsic linewidth of 0.85 kHz was deployed. This ensures that the laser's inherent phase-coherence time vastly exceeds the characteristic fluctuation timescales of the turbulence-induced random media, effectively eliminating source-induced phase noise from the environmental measurements. For coherent detection, an intermediate frequency (IF) of 80 MHz was chosen to provide high temporal resolution. The received signal was processed using a short-time Fourier transform (STFT) with a 10 kHz bin width—safely satisfying the Nyquist criterion required to capture the multiscale temporal dynamics of the atmospheric turbulence and ensuring high-fidelity signal reconstruction.

The localized space-time structure of the atmospheric random media was mapped in situ utilizing an array of TempVue20 Pt100 temperature sensors paired with a CR1000X datalogger. By systematically varying the spatial separation between sensors, the fluctuations of the squared refractive index were empirically derived via the relation $\mu(t,\boldsymbol{x}) \approx \left[ C_{3} p(t,\boldsymbol{x}) / T(t,\boldsymbol{x}) + C_{4} \right]^{2}$ \cite{mathar2007RefractiveIndexHumid}, where $p$ and $T$ denote the instantaneous atmospheric pressure and temperature, respectively. This precise micrometeorological data serves as the physical ground truth, enabling the direct computation of the empirical covariance function $K_{\mu}(t_{0},r) = \mathbb{E}[W(t_{0},\boldsymbol{0}) W(t_{0},\boldsymbol{r})]$ characterizing the media.

Optically, the system employs a heterodyne architecture. The laser output is partitioned via a 95:5 fiber coupler into signal and local oscillator (LO) paths. The signal beam is frequency-upshifted by 80 MHz using an acousto-optic modulator and expanded to a 300 mm diameter via a Galilean telescope to mitigate finite-aperture diffraction penalties. The beam then propagates along a 588-m near-ground outdoor atmospheric channel. Active tip-tilt correction is applied continuously to suppress global beam wander and isolate higher-order wavefront distortions. At the receiver terminus, a variable-aperture assembly couples the collected light into a balanced detector. The resulting 80 MHz IF beat signal is digitized by a 12-bit analog-to-digital converter and processed in real time.

\begin{figure}[htbp]
	\centering
	\includegraphics[height=8cm]{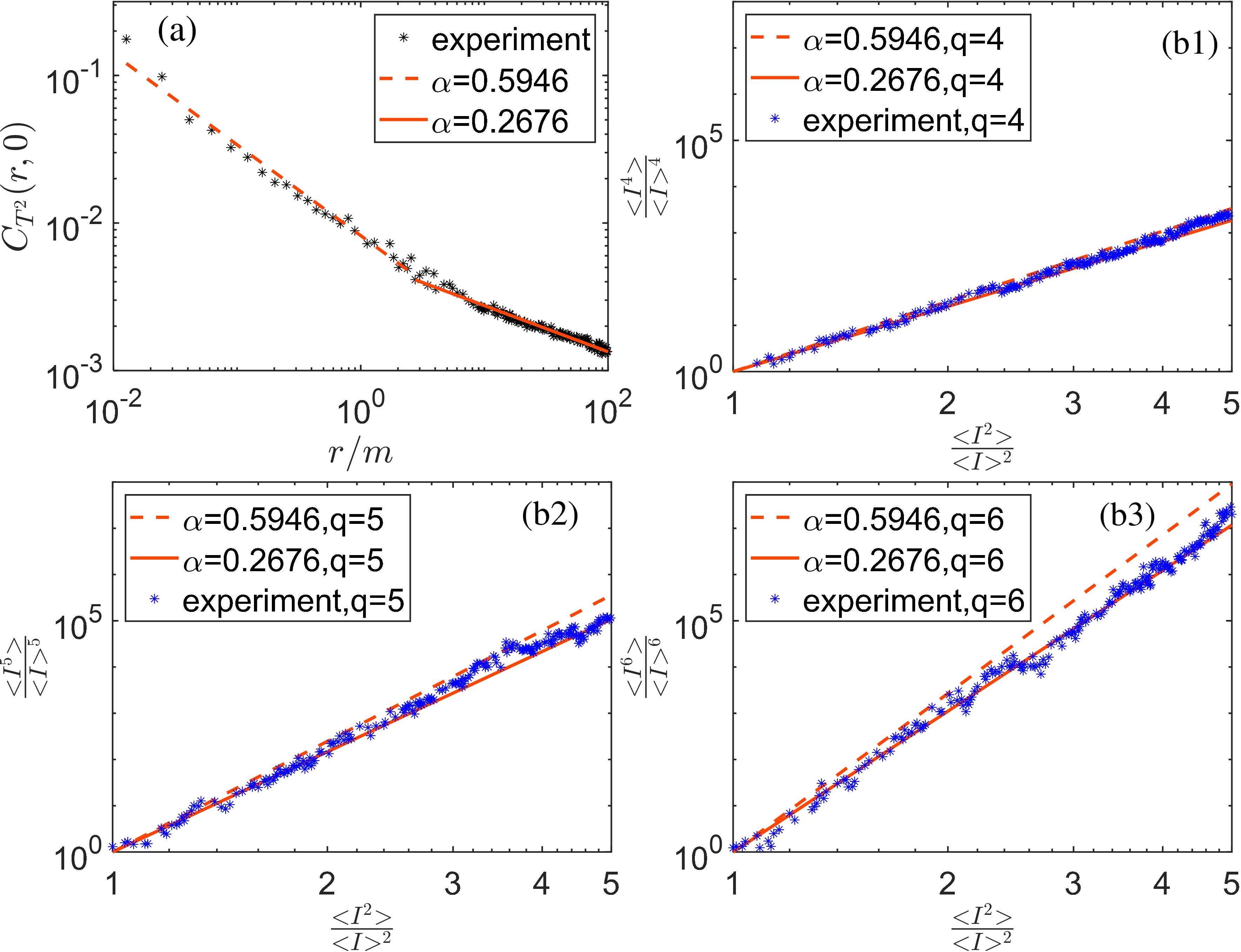}
	\caption{Space-time characterization of the random media and experimental validation of intensity moments. (a) Double-logarithmic plot of the squared refractive index fluctuations versus separation distance $r$, derived from outdoor measurements using the TempVue 20 Pt100 sensors. The experimental data (asterisks) are characterized by two piecewise linear fits (solid and dashed lines), highlighting the spatial scaling behavior. (b1)–(b3) Normalized higher-order intensity moments $\langle I^q \rangle / \langle I \rangle^q$ ($q = 4, 5, 6$) as a function of the second-order moment $\langle I^2 \rangle / \langle I \rangle^2$ for a point-like 2 mm receiving aperture. The striking agreement spanning multiple orders of magnitude confirms the analytical predictions of Eq.\eqref{w13}.}
	\label{F2}
\end{figure}

A specialized digital signal processing pipeline was developed to extract the instantaneous amplitude and phase of the propagated optical field. The raw, continuous data stream is temporally segmented into successive analysis blocks: $\{(0,t_{1}), (t_{1},t_{2}), \cdots, (t_{i},t_{i+1}), \cdots\}$. Fourier analysis within each block pinpoints the exact spectral peak of the IF carrier, allowing for the isolation of the corresponding complex vector. From this vector, the instantaneous optical field is mathematically reconstructed as $A_i e^{j\varphi_{i}}$, where $A_i$ and $\varphi_i$ are the amplitude and phase at the discrete time instance $t_i$. Crucially, independent baseline calibrations using extended optical fiber delay lines confirmed that any residual phase noise originating from the laser or the algorithmic partitioning was structurally negligible compared to the atmospheric signature. All subsequent intensity metrics and higher-order statistical moments were synthesized from these purified amplitude and phase time-series.

Following the acquisition of the experimental dataset, we first evaluated the scaling behavior of $\mu(t_0, \boldsymbol{x})$ at a fixed time snapshot $t_0$ across varying propagation distances $r$. As shown in Fig.\ref{F2}(a), the spatial power-law scaling was distinctly captured using a two-segment piecewise fit in double-logarithmic coordinates. To probe the local intensity statistics, the receiver aperture was initially restricted to a point-like 2 mm diameter, with the balanced detector signal sampled at 250 MHz. Figures \ref{F2}(b1)-(b3) illustrate the empirical dependence of the normalized higher-order moments $\dfrac{\langle I^{q} \rangle}{\langle I \rangle^{q}} \bigg\vert_{\boldsymbol{x}_0}$ against the second-order moment $\dfrac{\langle I^{2} \rangle}{\langle I \rangle^{2}} \bigg\vert_{\boldsymbol{x}_0}$ for $q=4,5,6$.

\begin{figure}[htbp]
	\centering
	\includegraphics[height=7cm]{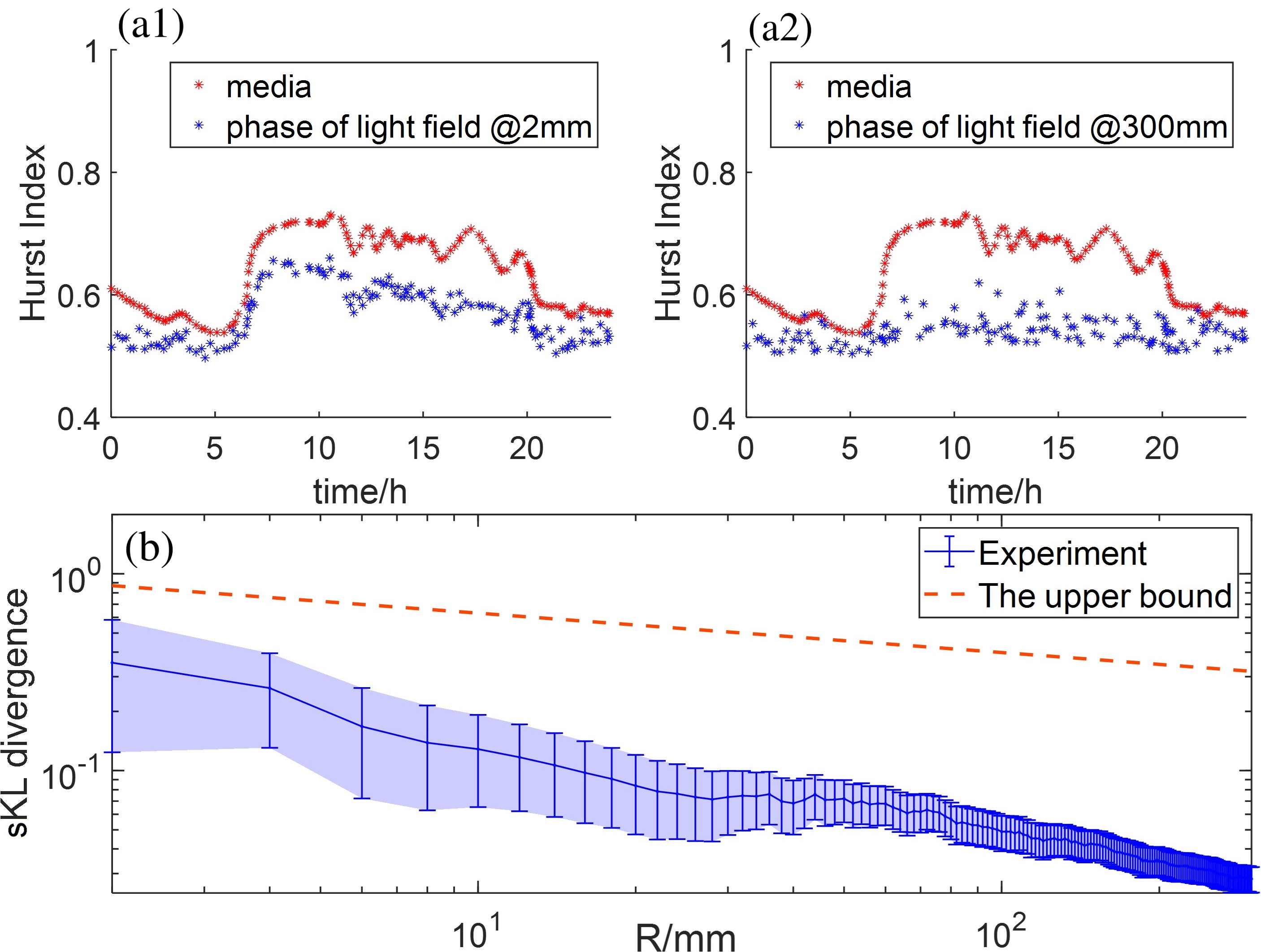}
	\caption{Non-Markovian temporal correlations and statistical convergence. (a1)-(a2) For a 2 mm aperture, the 24‑hour Hurst index variations of the environmental fluctuations $\mu(t_i, \boldsymbol{x}_0)$ and the optical phase $\varphi(t_i)$ are strongly correlated (Pearson coefficient 0.905), demonstrating memory transfer. For a 300 mm aperture, the correlation becomes insignificant (Pearson coefficient 0.341) due to spatial averaging. (b) The sKL divergence between the normalized process $Z_{R}(t)$ and the standard Gaussian $\mathcal{N}_{t} (0,1)$, plotted on a double‑logarithmic scale as a function of the aperture radius. The dashed line maps the theoretical upper bound attenuation $R^{-\frac{\alpha}{2}}$. These results empirically validate the predictions of Eq.\eqref{whurt} and Eq.\eqref{w15}.}
	\label{F3}
\end{figure}

To explicitly map the non-Markovian memory effects governing the wave dynamics, the receiver aperture was subsequently dilated from 2 mm up to the full 300 mm, while the data acquisition rate was maximized to 1 GHz. We utilized Rescaled Range (R/S) analysis \cite{mcleod1978PreservationRescaledAdjusted} (see Supplementary Information for explicit algorithmic implementations \cite{SeeSupplementalMaterialb}) to quantitatively estimate the Hurst index for both the environmental fluctuations and the optical phase variations:
\begin{align}
	&H_{X}=\lim_{n\rightarrow \infty}\frac{\log{R}/{S}(X_{1},\cdots,X_{n})}{\log n}, \; X_{i}=\{\mu(t_{i},\boldsymbol{x}_{0}),\varphi_{i}\}.
\end{align}
As depicted in Fig.\ref{F3}(a1)--(a2), for the small 2 mm aperture, the 24-hour variation of the Hurst index for the phase $\varphi(t_i)$ exhibits a profound positive correlation (Pearson coefficient $0.905$) with the environmental Hurst index of $\mu(t_i, \boldsymbol{x}_0)$, unambiguously demonstrating that the temporal persistence of the random media is imprinted onto the light field. Conversely, for the 300 mm aperture, this correlation effectively vanishes (Pearson coefficient $0.341$), indicating the suppression of local memory-driven fluctuations through spatial averaging.

Finally, we evaluated the statistical convergence of the spatial-averaged field toward a memoryless limit. Figure \ref{F3}(b) tracks the symmetrized Kullback-Leibler (sKL) divergence between the normalized experimental process $Z_R(t)$ and the standard complex Gaussian distribution $\mathcal{N}_t(0,1)$ as the aperture radius $R$ increases. The divergence decays monotonically, tightly conforming to the theoretical upper bound attenuation $C_{2} R^{-\frac{\alpha}{2}}$ prescribed by our model. In conclusion, the robust consistency between these multi-dimensional experimental measurements and the analytical SPDE predictions decisively confirms the validity of our exact theoretical mapping for describing wave propagation in non-Markovian media.

	\section{Conclusion}
	
	In conclusion, we have established a rigorous theoretical framework based on stochastic partial differential equations (SPDEs) that extends beyond the conventional Markov approximation to model wave propagation in space-time random media. By incorporating temporal fractional Brownian motion (fBm) alongside spatial multifractal power-law scaling, we demonstrate that the evolution of the propagating light field maps onto the hyperbolic Anderson model. This analytical approach explicitly captures the intrinsic non-Markovian dynamics of refractive index fluctuations—most notably, the long-range temporal memory effects and colored noise characteristics that memoryless paradigms typically neglect.
	
	Crucially, this mathematical mapping enables the direct derivation of memory-dependent statistical properties of the emergent light field. We have systematically corroborated these non-Markovian scaling signatures through a macroscopic outdoor experimental platform. By concurrently measuring localized environmental fluctuations and heterodyne optical phase dynamics, we provide empirical evidence of memory transfer from the complex medium to the propagating wave. Furthermore, our quantitative analysis of the Hurst index correlations and the spatially-averaged sKL divergence bounds bridges the SPDE theory with observable optical phenomena, elucidating the physical mechanisms behind scintillation saturation and spatial aperture averaging.
	
	Ultimately, our results offer a physical basis to clarify long-standing anomalies in atmospheric and oceanic propagation, particularly where standard white-noise approximations fail. Beyond these specific media, the non-Markovian approach developed here can serve as a general tool for analyzing complex wave–matter interactions. Given that temporal persistence and colored noise appear naturally across many physical systems, explicitly accounting for these environmental memory effects will be critical for emerging photonic applications. Practically, this framework can inform the design of more reliable free-space optical communication links, turbulence-resistant coherent imaging techniques, and precise remote sensing methods in dynamic environments.
	\section*{Funding}
	National Key Research and Development Program of China (2024YFF0505602).
	
	\section*{Acknowledgment}
	C.W. and S.H. conceived the idea. C.W., J.Q., S.L., and C.D. performed the experimental measurements. C.W. analyzed the data. All the authors discussed the results and contributed to the manuscript.
	
	\section*{Disclosures}
	The authors declare no competing interests.
	
	\section*{Data availability}
	Data underlying the results presented in this paper are not publicly available at this time but may be obtained from the authors upon reasonable request.

	\section*{Supplementary Information}
	Detailed mathematical proofs, expanded theoretical derivations, and extended experimental data analyses are available in the Supplementary Information.
	
	\bibliographystyle{naturemag}
	\bibliography{Propagation_nM}
	
\end{document}